# Programmable recirculating bricks mesh architecture for photonic neural networks

Jacek Gosciniak

*Institute of Microelectronics and Optoelectronics, Warsaw University of Technology,*

*Koszykowa 75 st., Warsaw 00-662, Poland*

Email: jacek.gosciniak@pw.edu.pl

**Abstract**

General-purpose programmable photonic processors are considered a crucial technology because they combine the ultra-high speed, massive bandwidth, and energy efficiency of light-based computing with the flexibility of software-defined hardware. Unlike application-specific photonic circuits (ASPICs) designed for one task, these processors use reconfigurable waveguide meshes to implement various functions—such as switching, filtering, or AI computation—on a single chip, allowing for rapid prototyping and versatile, on-demand hardware redefinition.

Here we report a recirculating "bricks" mesh architecture that can be easily implemented in photonic neural networks. It will be shown that a single programmable optical system is capable of performing various functions depending on the requirements. In particular, we will show that the same network, after being reprogrammed, can perform many different functions, ranging from a crossbar network to optical interference circuits with variable structures, which can then be subjected to Singular Value Decomposition. Furthermore, the "bricks" mesh serves as an excellent foundation for implementing a monitoring system capable of monitoring the power in each location of the circuit and subsequently self-calibrating and stabilizing the circuit using a feedback loop.

**Introduction**

Neural networks (NNs) are computing models inspired by the structure of a biological brain that are trained on input data to implement complex signal processing tasks [**1, 2**]. However, the energy required for training doubles every 5 months, necessitating the development of more energy-efficient hardware for NNs [**3**].

To address this problem, programmable photonic neural networks (PNNs) have been proposed [**4-9**] as a promising, scalable, and mass manufacturable integrated photonic hardware solution which significantly accelerates the most expensive, but also the most fundamental mathematical operation in a PNN: unitary matrix-vector multiplication (MVM). Such operations can be performed at the speed of light and with minimal power consumption, thus, represent a promising fully optical computing paradigm.

The main factor limiting the performance of programmable photonic circuits is the phase shifter that cannot be scaled down arbitrarily as it is subject to trade-off between loss, footprint, power, speed, and many other factors. One way to avoid this trade-off is to find an architecture that minimizes the average phase shift per element in a large circuit. Essentially, this means that each switch should be able to operate with only a fraction $O(1/N)$ of the phase shifters active at any given time [**10**]. The first solution to this problem relies on the implementation of a crossbar network architecture [**11-14**]. The second solution, in our opinion, is based on the implementation of recirculating networks [**7-9, 15**] in which signal transmission is not limited to a single direction, usually from left to right, as in feed-forward networks, but it can propagate in any direction, including backward. Thanks to these new capabilities, recirculating networks enable a significant reduction in the number of switches in the system while maintaining all signal processing capabilities.

Furthermore, one of the main challenges in photonic information processing, especially in a deterministic quantum information processing, is to realize key transformations on time scales shorter than decoherence time [**16**] and loss mechanism [**15, 17**]. This can be realized through an architecture-based approach using the recirculating photonic networks that minimize the duration of information processing tasks through dynamic control of multi-photon interactions. It has been shown that recirculating photonic networks may provide significant improvements in time and hardware efficiency relative to state-of-the-art architectures based on neural networks [**16**].

Among recirculating networks, one in particular—based on the so-called recirculating "bricks" mesh architecture (or shifted rectangular mesh architecture)—deserves special attention due to its significant similarity to feed-forward networks [**15, 17, 18**].

**Recirculating "bricks" mesh architecture**

The recirculating "bricks" mesh architecture consists of from 2 to 4 MZIs in a unit cell compared to 6 for a hexagonal mesh which result in a significant reduction in the optical path length and, thus, the propagation losses [**15**]. Simultaneously, it yields a more efficient 3-point interconnection scheme compared to a regular square mesh architecture where each unit cell is connected through 4 points. The ability for signal flow in any direction enables the implementation of both control loops and various types of filters, including both infinite impulse response (IIR) filters, which are based on ring resonators (RRs), and finite impulse response (FIR) filters, which are based on asymmetric MZIs. This significantly exceeds the capabilities of feed-forward networks, enabling the implementation of more complex systems [**15, 17, 18**].

The reconfiguration performance of the "bricks" mesh, defined as the number of filters with different frequency separation values for the RR-based filter, amounts to 11 for only 25 MZIs. In comparison, for a hexagonal and triangular mesh with the same number of MZIs, it was calculated at 9 and 6, respectively, while for a regular square mesh, it was calculated at 6 [**15**]. Simultaneously, the reconfiguration performance for the MZI-based filter arranged in the "bricks" mesh architecture exceeds 12 which is exactly the same as the performance of the hexagonal mesh. In comparison, for a triangular mesh it was calculated at 8, while for a regular square mesh at 6 [**15**]. As observed from above, the "bricks" mesh architecture doubles the performance compared to a regular square mesh architecture. This is yet another undeniable advantage of an architecture built on a "bricks" mesh.

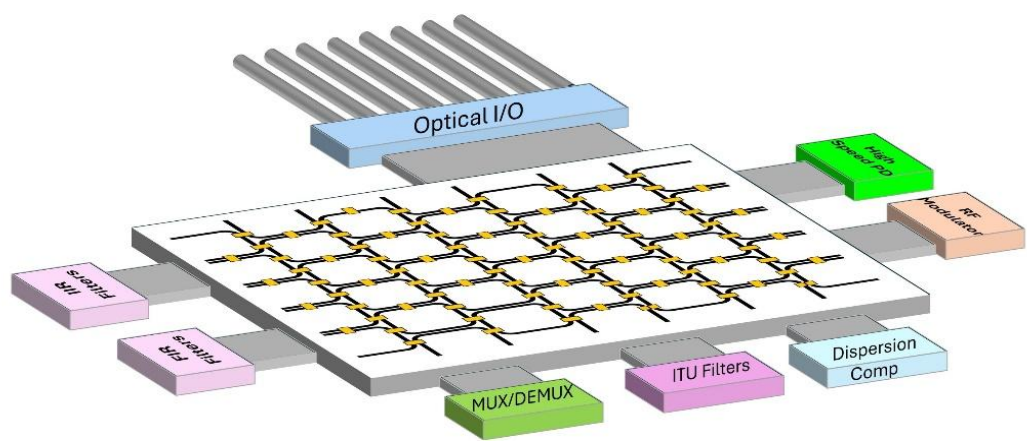

**Figure 1**. The photonic chip with a programmable mesh core connected to control electronics, optical fibers, filters, high-speed modulators and detectors and many other components.

Previous studies have shown [**17**] that the optical depth of a network based on the "bricks" mesh architecture is significantly reduced compared to feed-forward networks, both based on rectangular and triangular meshes, and the number of active MZIs required to perform any linear transformation between input and output modes is reduced by more than a factor of 13. Furthermore, all ports that can function as both input and output ports are located on all four sides of the mesh structure [**17**]. This enables the implementation of systems that are not achievable with feed-forward networks.

Simultaneously, however, the recirculating "bricks" mesh architecture closely resembles feed-forward networks, which enables the implementation of all tasks possible for this type of network. Additionally, the next advantage of this mesh is that it can be configured for various tasks such as, for example, the Singular Value Decomposition (SVD) and can be implemented in different types of network architectures such as, for example, the crossbar network.

**Crossbar architecture with recirculating "bricks" mesh**

One such network is a crossbar network [**11-14**]. Crossbar networks are one of the most exciting AI hardware architectures to construct a photonic matrix multiplier of a given number of input and output dimensions able to achieve parallelism and analog computing at the same time. It typically aligns incoming signals along one direction (i.e., left-right) and outgoing signals along the other direction (top-down). The crossbar network consists of $N$ input and $M$ output ports with $N \times M$ connections, allowing for all-to-all connectivity at the cost of $N \times M$ reconfigurable coupling elements.

In this paper, the crossbar network is constructed of the MZI mesh that uses collections of MZIs as a linear optical interference unit (OIU) that can perform calculations through their transfer matrices. An example provided in this paper consist of an $N \times M$ arbitrary unitary matrix that consists of MZIs arranged in recirculating "bricks" mesh topology. Compared to standard feed-forward mesh topologies where the total number of MZI units for $N \times N$ matrix is exactly $N(N-1)/2$, i.e., consist of $O(N^2)$ parameters (2 phase shifters) [**19, 20**], in the recirculating "bricks" mesh topology the total number of MZI units scales with $N$ which represents a significant reduction in the number of MZIs required to perform the relevant logical operations [**15**].

The output from each column $z_j$ is described by McCulloch-Pitts neuron model [**21**] and is expressed as

$$z_j = \sigma_i \left( \sum w_{ij} x_i \right)$$

where $z_j$ is the output vector from the *jth* line made of $M$ outputs, $\sigma$ is an activation (nonlinear) function, $x_i$ is the *ith* element of the input vector $X$ with $N$ elements and $w_{ij}$ is the $N \times M$ weight matrix for the input value $x_i$. The linear term $y_j = \sum w_{ij} x_i$ represents the weighted addition while the nonlinear part comprises the activation function $\sigma$.

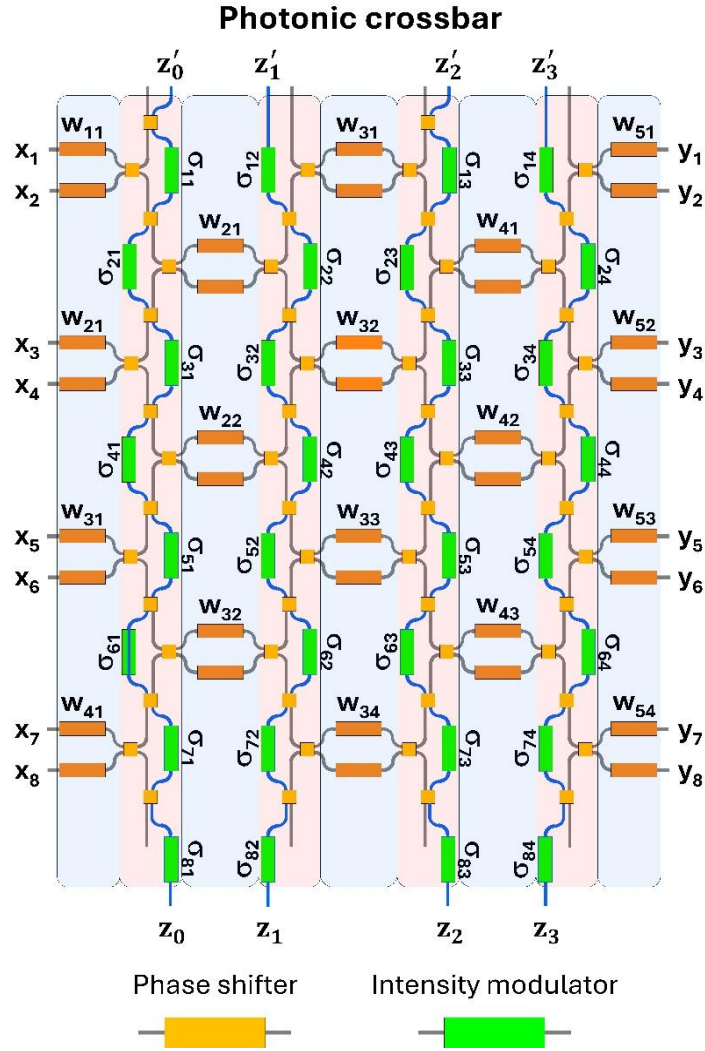

**Figure 2**. The proposed crossbar network based on a "bricks" mesh architecture.

The optical data signal vector $X = [x_1, x_2, ..., x_n]^T$ enters the crossbar array via *i-th* row and then each signal is weighted accordingly using phase shifters implemented in a symmetric MZI (modulated through a phase shifter) (**Fig. 2**). Each two adjacent signals $x_{n-1}$ and $x_n$ are then combined via a 3 dB beam splitter. Depending on the phase shifts on two signals, the resulting signal is split in vertical direction, up or down. In the next 3 dB splitter the optical signal is split again, so that the part of the column signal gets coupled to the row route while the remaining part of the column signal continues its vertical route along the column until reaching the next beam splitter. The columns consist of intensity modulators which are responsible for signal activation. After the next two beam splitters the signal will be split again into a portion entering the 2$^{nd}$ row and a portion that gets forwarded to its 3$^{rd}$ row. Each time a signal is routed to the appropriate column, it is activated using an activation function. This process is repeated multiple times until reaching the final column. The output signals $z_j$ at each column correspond to the inner product between the input vector $x_i$ and the respective weights $w_{ij}$. Thus, in each column a signal is weighted, summated and then activated.

Depending on the requirements, all activation function units can be deactivated, with the exception of the last ones in each column. Deactivation at this point applies to the settings of the intensity modulators corresponding to maximum signal transmission. Consequently, only the last intensity modulators in each column can function as active activation function units (**Fig. 2**). In this way, all signals reaching a given column from successive rows will be summed, and only after being completely summed they will be subjected to the activation function.

In other scenario (**Fig. 2**) the activation function is activated every time an additional signal is added to the output column signal. Thus, the first weighted addition is followed by a nonlinear activation, after which another weighted addition occurs in response to the next signal, followed by another nonlinear activation. This happens every time a signal from the successive row is added to the output signal $z$.

**Singular Value Decomposition with recirculating "bricks" mesh**

As can be seen from the above discussion, the photonic neural network (PNN) is composed of two essential blocks: an optical interference unit (OIU) that implements optical matrix multiplication and separated nonlinear optical function unit (NOFU) that implements the nonlinear activation [**22**]. The OIU provides an opportunity to process optical signals, which contain information in amplitude and phase, directly within network while bypassing slow optical-to-electronic conversion. The linear transformations are realized by computing matrix-vector products through an interference in a MZI mesh that consist of $N$ input and output mode. The OIU consists of a cascade of $N$ layers, such as binary tree layers (**Fig. 3**) or diagonal lines (**Fig. 4**) (resulting in a triangular mesh), all constructed from $2 \times 2$ programmable interferometer blocks implemented using integrated MZIs. The fundamental function of the MZI is to be able to minimize to zero a power in either of its output powers given any input vector. These layers may then learn a decomposition of the corresponding transformation matrix via a sequential, layer-by-layer power optimization method relying on measurement and feedback [**18, 22, 23**].

In this arrangement, double unitary matrix modulations and a diagonal matrix modulation are deployed for arbitrary matrix modulation since an arbitrary matrix could be decomposed to the multiplication of a unitary matrix, a diagonal matrix, and another unitary matrix through singular value decomposition (SVD).

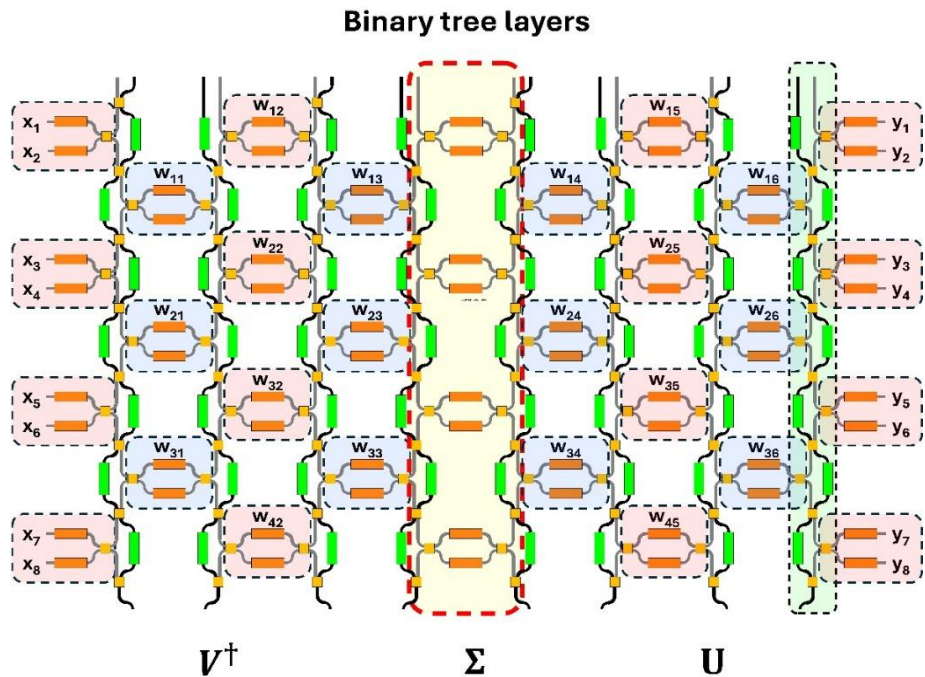

**Figure 3**. Binary tree layers of Mach-Zehnder interferometers (MZIs) that may serve for layer-by-layer power optimization.

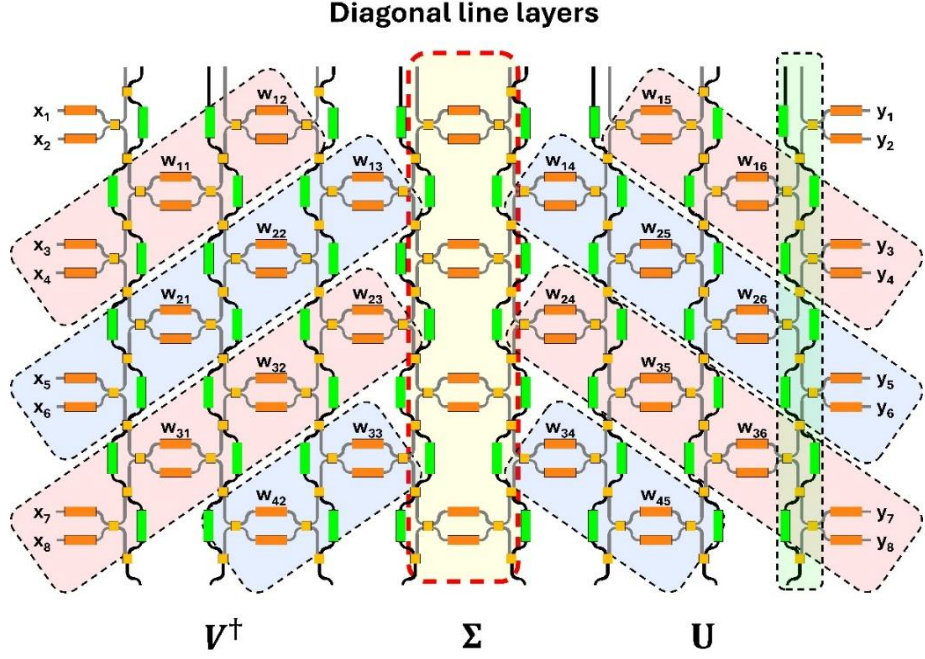

**Figure 4**. Diagonal line layers of Mach-Zehnder interferometers (MZIs) that may serve for layer-by-layer power optimization.

Such full MZI is made from two phase shifters $\phi_1$, $\phi_2$ and two 50:50 beam splitters. The 50:50 beam splitter can be represented by a matrix

$$U_{BS} = \frac{1}{\sqrt{2}}\begin{bmatrix} 1 & i \\ i & 1 \end{bmatrix} \quad (1)$$

while the phase-shifter by a matrix

$$U_{PS} = \begin{bmatrix} e^{i\phi} & 0 \\ 0 & 1 \end{bmatrix} \quad (2)$$

Thus, the transfer function of the full symmetric MZI (sMZI) (BS-MZI-BS) is expressed as

$$U_{sMZI} = U_{BS}\begin{bmatrix} e^{i\phi_1} & 0 \\ 0 & e^{i\phi_2} \end{bmatrix}U_{BS} = \frac{1}{2}\begin{bmatrix} e^{i\phi_1} - e^{i\phi_2} & i(e^{i\phi_1} + e^{i\phi_2}) \\ i(e^{i\phi_1} + e^{i\phi_2}) & -e^{i\phi_1} + e^{i\phi_2} \end{bmatrix}$$

$$= ie^{i\left(\frac{\phi_1+\phi_2}{2}\right)}\begin{bmatrix} \sin\left(\frac{\phi_1-\phi_2}{2}\right) & \cos\left(\frac{\phi_1-\phi_2}{2}\right) \\ \cos\left(\frac{\phi_1-\phi_2}{2}\right) & -\sin\left(\frac{\phi_1-\phi_2}{2}\right) \end{bmatrix} \quad (3)$$

In comparison, the transfer function of the full asymmetric MZI (aMZI) (BS-MZI-BS-MZI) is expressed as

$$U_{aMZI} = \frac{1}{2}U_{BS}\begin{bmatrix} e^{i\phi_2} & 0 \\ 0 & 1 \end{bmatrix}U_{BS}\begin{bmatrix} e^{i\phi_1} & 0 \\ 0 & 1 \end{bmatrix} = ie^{i\frac{\phi_2}{2}}\begin{bmatrix} e^{i\phi_1}\sin\left(\frac{\phi_2}{2}\right) & \cos\left(\frac{\phi_2}{2}\right) \\ e^{i\phi_1}\cos\left(\frac{\phi_2}{2}\right) & -\sin\left(\frac{\phi_2}{2}\right) \end{bmatrix} \quad (4)$$

Here, the MZI is implemented with two internal and no external phase shifters, thus, it is called a symmetric MZI (sMZI). This type of sMZI is attractive because it is more compact without the need for an external phase-shifter, which can account for a significant fraction of the length of the circuit. A shorter structure not only occupies less area on a chip but also suffers from less propagation losses what is especially important in photonic neural networks (PNNs).

The OIU can implement any real-valued matrix $W$ that may be decomposed as $W = U\Sigma V^\dagger$ through singular value decomposition (SVD), where $U$ is a unitary matrix, $\Sigma$ is a rectangular diagonal matrix with non-negative real numbers on the diagonal and $V^\dagger$ is the complex conjugate of the unitary matrix $V$. As it has been previously shown, any unitary transformations $U$ and $V^\dagger$ can be implemented with phase shifters and optical beam splitters. In comparison, $\Sigma$ can be implemented using optical attenuators [24]. In consequence, the implementation of matrix multiplication with unitary matrices consumes no power [22], which enables the extreme energy efficiency of the ONN architectures, especially taking into account that major proportion of ANN calculations involve matrix products [22].

In a standard photonic arrangement, the NOFU can be implemented through an additional photonic layer [22], however, in a proposed recirculating "bricks" mesh architecture, the NOFU can be implemented through any vertical lines as shown in **Fig. 3** and **4** (shadowed green region) using optical nonlinearities such as saturable absorption [4] and bistability [25]. This is an undeniable advantage of the proposed system, as it eliminates the need to implement additional layers, thereby reducing the system's footprint. Additionally, the system is programmable in real time and can perform a variety of functions depending on current needs [26-28].

An operation principle for those two arrangements is similar to a previously discussed crossbar network. The optical signal vector $X = [x_1, x_2, \ldots, x_n]^T$ is splitted equally between all MZI modulators each of which encode an element of $x_i$ into amplitude and phase. Then, a programmable MZI mesh performs transformation

$$y_i^1 = W^1 x_i^1$$

through optical interference. At the end of layer, the signals enter the programmable NOFU that applies the activation function $\sigma$ to yield the input to the next layer. As previously noted, NOFU can be implemented as an additional layer separating successive OIUs, or through the vertical elements shown in **Fig. 3** and **4** (shadowed green area). Depending on the system's characteristics, a PNN may consist of a large number of both OIUs and NOFUs which may implement additional transformations

$$y_i^n = \sigma(W^n x_i^n)$$

As observed, interference and thus processing take place entirely in the optical domain without assistance of electronics.

**Self-configuration and stabilization**

Another advantage of a proposed recirculating "bricks" mesh architecture is its compatibility with a recently proposed monitoring system that can monitor a signal flow in each part of the circuit under operation [18]. Furthermore, it can perform a control strategy to automatically adjust the optical power and phase of photonic components at each point of the photonic system with extreme accuracy and minimal insertion loss. The technique is based on a feedback control loop that simultaneously adjusts the matrix coefficient of the device transfer function and compensates for process tolerances and thermal drift in real time. This control strategy relies on the Wheatstone bridge arrangement with a calibration-free feedback loop that does not require prior knowledge of the device transfer function [18].

As all bridge components are located on the same chip in close proximity to each other, the system exhibits minimal sensitivity to changes in the temperature of the photonic components situated on the chip. Therefore, the only factor affecting the temperature change of the component under test is the optical power propagating through this section of the waveguide. Furthermore, unlike most monitors and photodetectors, which provide an electrical current upon measuring an optical signal, the proposed arrangement provides a voltage [**18**]. This eliminates the need for current-to-voltage conversion at a later stage, thereby streamlining the circuit and substantially reducing noise level.

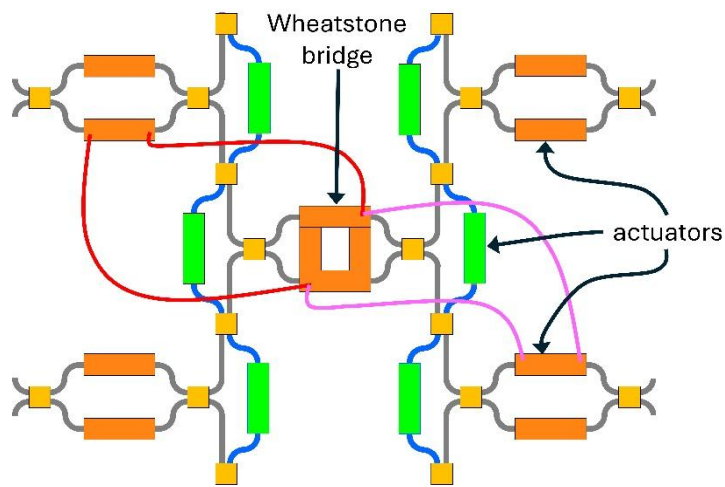

**Figure 5**. A schematic view of the local feedback with the monitoring module and with the actuators placed either in a preceding or succeeding part of the system.

The system can be self-calibrated using feedback loops with voltage applied to both preceding and succeeding actuators, depending on the requirements, to stabilize the circuit at the desired value defined by the target voltage in a processing unit [**Fig. 5**] [**18**]. Furthermore, the monitoring system can be implemented to self-configure entire layers, for example binary tree layers and/or diagonal lines [**23**], as presented in **Fig. 3 and 4**, which significantly expands the range of available design arrangements.

The system can be built based on the transparent conductive oxides (TCOs) that are characterized by the epsilon-near-zero (ENZ) point whose properties can be actively modified by the external voltage [**29-31**].

**Summary**

Recirculating meshes, and in particular "bricks" mesh, are capable of providing significant benefits in many key technologies, including photonic neural networks, primarily due to their scalability, high reconfigurability, and lower optical losses compared to traditional, large-scale static photonic circuits. Instead of building a massive, one-way interferometer, a recirculating mesh allows photons to pass through a smaller, programmable, and tunable component multiple times to simulate a larger, complex unitary transformation. This enables the realization of key transformations on time scales shorter than the decoherence time and the loss mechanism. Furthermore, they are capable of minimizing the duration of information processing tasks through dynamic control of multi-photon interactions. It has been shown that such networks may provide significant improvements in time and hardware efficiency relative to state-of-the-art architectures based on neural networks. These advantages make "bricks" mesh designs a very promising technology for implementing practical, near-term, large-scale quantum signal processing and neural networks.

**Acknowledgement**

The author is very thankful to Prof. D. G. Misiek for his support and very valuable suggestions.